\documentclass[aps,prl,twocolumn,showpacs,preprintnumbers,amsmath,amssymb]{revtex4-1}

 \def\ep{{\epsilon}}

 \def\frac#1#2{{#1\over #2}}

 \def\s{\sqrt}

\def\be{\begin{equation}}
\def\ee{\end{equation}}
\def\ba{\begin{eqnarray}}
\def\ea{\end{eqnarray}}

\def\({\left(}
\def\){\right)}

 \def\f {\frac}

 \def\ddd{\cdot\cdot\cdot}
 \def\no{\nonumber \\}

 \def\la{\langle}
 \def\lb{\rangle}
 \def\ep{\epsilon}

\usepackage{color}
\usepackage{graphicx}
\usepackage{dcolumn}
\usepackage{bm}

\begin{document}

\title{Quantum Dimension as Entanglement Entropy in 2D CFTs}
YITP-14-17; IPMU14-0046
\author{Song He$^{a,b}$, Tokiro Numasawa$^a$,  Tadashi Takayanagi$^{a,c}$ and Kento Watanabe$^a$}

\affiliation{$^a$Yukawa Institute for Theoretical Physics,
Kyoto University, \\
Kitashirakawa Oiwakecho, Sakyo-ku, Kyoto 606-8502, Japan}

\affiliation{$^{b}$State Key Laboratory of Theoretical Physics, Institute of Theoretical Physics, \\ Chinese Academy of Science, Beijing 100190, P. R. China}

\affiliation{$^{c}$Kavli Institute for the Physics and Mathematics
 of the Universe,\\
University of Tokyo, Kashiwa, Chiba 277-8582, Japan}

\date{\today}

\begin{abstract}
We study entanglement entropy of excited states in two dimensional conformal
field theories (CFTs). Especially we consider excited states obtained by acting primary operators on a vacuum. We show that under its time evolution, entanglement entropy increases by a finite constant when the causality condition is satisfied. Moreover, in rational CFTs, we prove that this increased amount of (both Renyi and von-Neumann) entanglement entropy always coincides with the log of quantum dimension of the primary operator.

\end{abstract}

\maketitle

\section{Introduction}

Quantum field theories (QFTs) contain infinitely many degrees of freedom and therefore we can define so many kinds of observables in general.
Among them, entanglement entropy is a very helpful quantity especially when we would like to study global structures of any given quantum field theory. It is defined as the von-Neumann entropy $S_A=-\mbox{Tr}[\rho_A\log \rho_A]$ of the reduced density matrix $\rho_A$
for a subsystem $A$. The reduced density matrix $\rho_A$ is defined from the original density matrix $\rho$ by tracing out the subsystem $B$ which is the complement of $A$. For example, we can quantify topological properties by computing topological contributions in entanglement entropy, so called topological entanglement entropy \cite{wen}.

One may wonder if there is a sort of topological contribution in entanglement entropy even for gapless theories, especially conformal field theories (CFTs). The main aim of this paper is to extract such a quasi-topological quantity from (both Renyi and von-Neumann) entanglement entropy of excited states in two dimensional rational CFTs. Refer to an earlier work \cite{FFN} for a connection between the topological entanglement entropy and boundary entropy and to \cite{CC} for the one between the boundary entropy and entanglement entropy.

The $n$-th Renyi entanglement entropy is defined by
\be
S^{(n)}_A=\f{1}{1-n}\log\mbox{Tr}[\rho_A^n].
\ee
The limit $n\to 1$ coincides with the (von-Neumann) entanglement entropy.
We are interested in the difference of $S^{(n)}_A$ between the excited state and the ground state, denoted by $\Delta S^{(n)}_A$. Replica calculations
of $\Delta S^{(n)}_A$ for excited states defined by local operators have been formulated in \cite{UAM,NNT,Nozaki}. In particular, we will closely follow the construction in \cite{NNT}, which can be applied to QFTs in any dimensions. More details can be found in \cite{Nozaki}. Indeed, this quantity is topological as the late time values of $\Delta S^{(n)}_A$ do not change under any smooth deformations of subsystem $A$ \cite{NNT}. Also as we will see, this quantity gets non-trivial owing to a global structure of conformal blocks of CFTs.

 Consider an excited state which is defined by acting a primary operator $\mathcal{O}_a$ on the vacuum $|0\lb$ in a two dimensional CFT. We employ the Euclidean formulation and introduce the complex coordinate $(w,\bar{w})=(x+i\tau,x-i\tau)$ on $R^2$ such that $\tau$ and $x$ are the Euclidean time and the space, respectively. We insert the operator $\mathcal{O}_a$ at $x=-l<0$ and consider its real time-evolution from time $0$ to $t$ under the Hamiltonian $H$. This corresponds to the following density matrix:
\ba
\rho(t)&=&{\mathcal N}\cdot e^{-iHt}e^{-\ep H}\mathcal{O}_a(-l)|0\lb\la 0|{\mathcal{O}}^{\dagger}_a(-l)e^{-\ep H}e^{iHt} \no
&=& {\mathcal N}\cdot\mathcal{O}_a(w_2,\bar{w_2})|0\lb\la 0|
{\mathcal{O}}^{\dagger}_a(w_1,\bar{w}_1),
\ea
where ${\mathcal N}$ is fixed by requiring Tr$\rho(t)=1$.
Here we defined
\ba
&& w_1=i(\epsilon -it)-l, \ \ w_2 = -i(\epsilon+it)-l,   \label{wco} \\
&& \bar{w}_1=-i(\ep-it)-l,\ \ \bar{w}_2=i(\epsilon+it)-l.  \label{wcor}
\ea
An infinitesimal positive parameter $\ep$ is an ultraviolet regularization and we treat
$\ep\pm it$ as purely real numbers until the end of calculations as in \cite{cag,NNT,Nozaki}.

 \begin{figure}
  \centering
  \includegraphics[width=7cm]{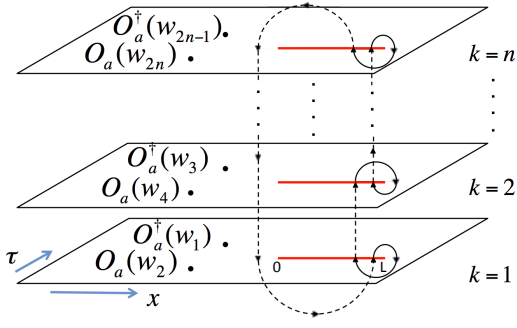}
  \caption{The $n$-sheeted space $\Sigma_n$. The red interval describes the subsystem A.}\label{rep}
\end{figure}

To calculate $\Delta S^{(n)}_A$, we employ the replica method  in the path-integral formalism
by generalizing the formulation for ground states \cite{CC} to our excited states \cite{NNT}.
We choose the subsystem $A$ to be an interval $0\leq x\leq L$ at $\tau=0$.
It leads to a $n$-sheeted Riemann surface $\Sigma_n$ with $2n$ operators ${\mathcal O}_a$ inserted
as in Fig.\ref{rep}.
In the end, we find that $\Delta S_A^{(n)}$ can be computed as
\ba
&&\Delta S_A^{(n)} \no
&&=\!\f{1}{1-n} \Biggl[\log{\left\langle{\mathcal{O}}^{\dagger}_a(w_l,\bar{w}_1)\mathcal{O}_a(w_2,\bar{w}_2)
\!\ddd\! \mathcal{O}_a(w_{2n},\bar{w}_{2n})\right\rangle_{\Sigma_n}}\no
&& \ \ \ \ -n\log\left\langle{\mathcal{O}}^{\dagger}_a(w_l,\bar{w}_1)
\mathcal{O}_a(w_2,\bar{w}_2)\right\rangle_{\Sigma_1}\Biggr], \label{replica}
\ea
where $(w_{2k+1},w_{2k+2})$ for $k=1,2,\ddd,n-1$ are $n-1$ replicas of $(w_1,w_2)$ in the $k$-th sheet of $\Sigma_n$. The term in the second line is given by a $2n$ points correlation function on $\Sigma_n$. The final term is a two point function on $\Sigma_1=R^2$ and we normalized this such that
\ba
\la \mathcal{O}^{\dagger}_a(w_1,\bar{w}_1)\mathcal{O}_a(w_2,\bar{w}_2)\lb_{\Sigma_1}
=\f{1}{|w_{12}|^{4\Delta_a}}=\f{1}{(2\ep)^{4\Delta_a}}, \label{ntw}
\ea
which is equal to ${\mathcal N}^{-1}$. Here $\Delta_a$ is the (chiral and anti-chiral) conformal dimension of the operator $\mathcal{O}_a$.

\section{Renyi Entanglement Entropy for $n=2$}

Let us first study the $n=2$ case $\Delta S^{(2)}_A$ in detail. Later we will generalize the results to any $n$. We can apply the conformal transformation:
\be
\frac{w}{w - L} = z^n , \label{cmap}
\ee
which maps $\Sigma_n$ to $\Sigma_1$. Setting $n=2$ and using (\ref{wco}), the coordinates $z_i$ are given by
(similarly $\bar{z}_i$ using (\ref{wcor}))
\ba
&& z_1 = -z_3 = \sqrt{ \frac{l - t - i \ep }{l + L - t - i \ep } } , \no
&& z_2 = -z_4 = \sqrt{ \frac{l - t + i \ep }{l + L - t + i \ep } } . 
\ea
It is useful to define the cross ratios $(z , \bar{z})$
\ba
z = \f{z_{12}z_{34}}{z_{13}z_{24}},\ \ \  \bar{z} = \f{\bar{z}_{12}\bar{z}_{34}}{\bar{z}_{13}\bar{z}_{24}},
\ea
where $z_{i j} = z_i - z_j$.  We would like to study the behavior of $(z , \bar{z})$  in the limit $\ep\to 0$. When $0<t<l$ or $t>L+l$, we find $(z,\bar{z})\to(0,0)$:
\be
z\simeq\f{L^2\ep^2}{4(l-t)^2(L+l-t)^2},\ \  \bar{z}\simeq\f{L^2\ep^2}{4(l+t)^2(L+l+t)^2}.\nonumber
\ee
In the other case $l<t<L+l$, we find $(z,\bar{z})\to(1,0)$:
\be
z\simeq 1-\f{L^2\ep^2}{4(l-t)^2(L+l-t)^2},\ \  \bar{z}\simeq \f{L^2\ep^2}{4(l+t)^2(L+l+t)^2}.\nonumber
\ee
 Though this limit $(z,\bar{z})\to(1,0)$ does not seem to respect the complex conjugate, it inevitably arises via our analytical continuation of $t$ from imaginary to real values.

Owing to the conformal symmetry, the four point function on $\Sigma_1$ can be expressed as
\ba
&& \la {\mathcal O}^{\dagger}_a(z_1,\bar{z}_1){\mathcal O}_a(z_2,\bar{z}_2){\mathcal O}^{\dagger}_a(z_3,\bar{z}_3){\mathcal O}_a(z_4,\bar{z}_4)\lb_{\Sigma_1} \no
&& =|z_{13}z_{24}|^{-4\Delta_a}\cdot G_a(z,\bar{z}). 
\ea
Applying the conformal map (\ref{cmap}), we obtain the four point function on $\Sigma_2$:
\ba
&& \la {\mathcal O}^{\dagger}_a(w_1,\bar{w}_1){\mathcal O}_a(w_2,\bar{w}_2){\mathcal O}^{\dagger}_a(w_3,\bar{w}_3){\mathcal O}_a(w_4,\bar{w}_4)\lb_{\Sigma_n}\no
&& =\prod_{i=1}^{4}\left|\f{dw_i}{dz_i}\right|^{-2 \Delta }
\la {\mathcal O}^{\dagger}_a(z_1,\bar{z}_1){\mathcal O}_a(z_2,\bar{z}_2){\mathcal O}^{\dagger}_a(z_3,\bar{z}_3){\mathcal O}_a(z_4,\bar{z}_4)\lb_{\Sigma_1} \no
&& = (4L)^{-8\Delta_a} \left| \f{(z_1^2 - 1) (z_2^2 - 1)}{z_1z_2} \right|^{8\Delta_a} \cdot G_a(z,\bar{z}).
\ea
Using this and (\ref{ntw}), the relevant ratio is expressed as a function which depends only on $z$:
\ba
&&  \f{\la {\mathcal O}^{\dagger}_a(w_1,\bar{w}_1){\mathcal O}_a(w_2,\bar{w}_2){\mathcal O}^{\dagger}_a(w_3,\bar{w}_3){\mathcal O}_a(w_4,\bar{w}_4)\lb_{\Sigma_n}}{\left(\la {\mathcal O}^{\dagger}_a(w_1,\bar{w}_1){\mathcal O}_a(w_2,\bar{w}_2)\lb_{\Sigma_1}\right)^2} \no
&& =|z|^{4\Delta_a}|1-z|^{4\Delta_a} \cdot G_a(z,\bar{z}) . \label{fgf}
\ea

\begin{figure}[htbp]
  \begin{center}
   \includegraphics[width=60mm]{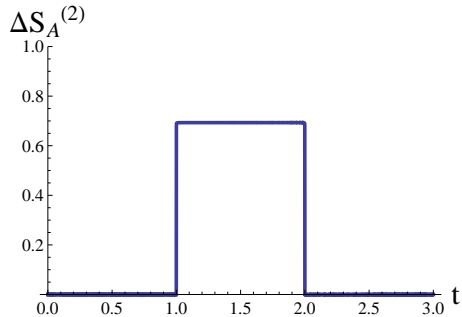}
  \end{center}
  \caption{The time evolution of $\Delta S^{(2)}_A$ for ${\mathcal O}_2$. We set $l = 1, L = 1$.}
  \label{fig:SA2}
\end{figure}

For example, let us consider a $c=1$ CFT defined by a (non-compact) massless free scalar $\phi$ and choose two operators
\be
{\mathcal O}_1 = e^{\f{i}{2} \phi},\ \ \ {\mathcal O}_2=\f{1}{\s{2}}(e^{\f{i}{2} \phi} + e^{-\f{i}{2} \phi}), \label{opew}
\ee
which have the same conformal dimension $\Delta_1=\Delta_2 = \f{1}{8}$.
Then, the function $G_a(z , \bar{z})$ is found to be
\ba
&& G_1(z , \bar{z}) = \f{1}{ \sqrt{ \left| z \right| \left| 1 - z \right| } } ,\no
&& G_2(z , \bar{z})=\f{1}{2 \s{|z||1-z|}} \left( |z| + 1 + |1 - z| \right) .
\ea
It is obvious that the Renyi entropy always becomes trivial $\Delta S^{(2)}_A=0$ for the operator ${\mathcal O}_1$. For ${\mathcal O}_2$, we find
\be
\Delta S^{(2)}_A =
\left\{
\begin{array}{l}
 \  0  \ \ \ \ \ \ \ \ \left(0 <  t < l, \ \mbox{or}\  \  t> l + L   \right) , \\
 \ \log 2 \ \ \ \ \ \ \ \ \left(l < t  < l + L  \right) .
\end{array}
\right  . \label{scak}
\ee
This is depicted in Fig.\ref{fig:SA2}.

The reason why we find the trivial result for ${\mathcal O}_1$ is because the excited state $e^{\f{i}{2}\phi}|0\lb$ can be regarded as a direct product state $e^{\f{i}{2}\phi_L}|0\lb_L\otimes e^{\f{i}{2}\phi_R}|0\lb_R$ in the left-moving (L: chiral) and right-moving (R: anti-chiral) sector \cite{NNT}. Therefore it is not an entangled state.

On the other hand, ${\mathcal O}_2$ creates a maximally entangled state (or equally Einstein-Podolsky-Rosen  state):  $\f{1}{\s{2}}\left(e^{\f{i}{2}\phi_L}|0\lb_L\otimes e^{\f{i}{2}\phi_R}|0\lb_R+
e^{-\f{i}{2}\phi_L}|0\lb_L\otimes e^{-\f{i}{2}\phi_R}|0\lb_R\right)$, which carries the Renyi entropy $\log 2$ for any $n$ \cite{NNT}. At $t=0$, we insert this operator at the point $x=-l$, which creates an entangled pair. The pair propagates in the left and right directions at the speed of light. When $l<t<l+L$, one fragment stays on the subsystem $A$ and the other on $B$, which leads to the $\log 2$ entropy. When $0<t<l$ or $t>l+L$, both fragments live in $B$ and thus the entropy vanishes. This argument based on the causal propagations explains the result (\ref{scak}).

This behavior is universal for any primary operators in any CFTs as is clear from (\ref{fgf}), though the explicit value of Renyi
entropy for $l<t<l+L$ depends on the choice of operator and CFT as we will study below.

In general CFTs, the function $G(z,\bar{z})$ can be expressed using the conformal blocks
\cite{BPZ}:
\be
G_a(z,\bar{z})=\sum_{b}(C^{b}_{aa})^2F_a(b|z)\bar{F}_a(b|\bar{z}),
\ee
where $b$ runs over all primary fields. In our normalization, the conformal block
$F_a(b|z)$ behaves in the $z\to 0$ limit:
\be
F_a(b|z)=z^{\Delta_b-2\Delta_a}(1+O(z)),
\ee
$\Delta_b$ is the conformal dimension of ${\mathcal O}_b$.

Since we found $(z,\bar{z})\to (0,0)$ when $0<t<l$ or $t>l+L$, we get the behavior $G_a(z,\bar{z})\simeq |z|^{-4\Delta_a}$, as the dominant contribution arises when $b=0$ i.e. when ${\mathcal O}_b$ coincides with the identity ${\mathcal O}_0(\equiv I)$.
Applying (\ref{fgf}), we get $\Delta S^{(2)}_A=0$, as expected from the causality argument.

 To analyze the entropy when the causality condition $l<t<l+L$ is satisfied, we need to apply the fusion transformation, which exchanges $z_2$ with $z_4$ (or equally $z$ with $1-z$):
\be
F_a(b|1-z)=\sum_{c}F_{bc}[a]\cdot F_a(c|z), \label{Ftr}
\ee
where $F_{bc}[a]$ is a constant, called Fusion matrix \cite{MSP,MS}.
In the limit $(z,\bar{z})\to (1,0)$, we obtain
\be
G_a(z,\bar{z})\simeq F_{00}[a]\cdot (1-z)^{-2\Delta_a}\bar{z}^{-2\Delta_a}.
\ee
Therefore we find the following expression from (\ref{fgf}):
\be
\Delta S^{(2)}_A=-\log F_{00}[a].
\ee

 Moreover, in rational CFTs, based on the arguments of bootstrap relations of correlations functions \cite{MSP,Ver}, it was shown in \cite{MS} that $F_{00}[a]$ coincides with the inverse of the quantity called quantum dimension $d_a$:
\be
F_{00}[a]=\f{1}{d_a}=\f{S_{00}}{S_{0a}}, \label{qdm}
\ee
where $S_{ab}$ is the modular $S$ matrix of the rational CFT we consider. In this way we obtain the remarkably simple result for two dimensional rational CFTs:
\be
\Delta S^{(2)}_A=\log d_a,
\ee
when $l<t<l+L$.

For example, if we consider the $(p+1,p)$ unitary minimal model and choose ${\mathcal O}_{a}$ to be the $(m,n)$ primary operator \cite{BPZ}, we can explicitly confirm (\ref{Ftr}) and (\ref{qdm}) using the expressions of four point functions in \cite{DF} and
$\Delta S^{(2)}_A$ for $l<t<l+L$ is found to be
\be
\Delta S^{(2)}_A=\log \left[\f{(-1)^{n+m}\cdot\sin\left(\f{\pi (p+1)m}{p}\right)\sin\left(\f{\pi pn}{p+1}\right)}{\sin\left(\f{\pi (p+1)}{p}\right)\sin\left(\f{\pi p}{p+1}\right)}\right].
\ee

\section{Renyi Entropy for General $n$}

The $n$-th Renyi entanglement entropy can be obtained from the formula (\ref{replica}) by computing the $2n$ point functions. Owing to the previous discussions, since we are interested in the non-trivial time period: $l<t<l+L$, we can assume the limit $L\to \infty$ and employ the simple conformal map $w=z^n$. Then the $2n$ points $z_1, z_2,\ddd,z_n$ in the $z$ coordinate are given by
\ba
&& z_{2k+1}=e^{2\pi i\f{k}{n}}(i\ep+t-l)^{\f{1}{n}}=e^{2\pi i\f{k+1/2}{n}}(l-t-i\ep)^{\f{1}{n}} \no
&& z_{2k+2}=e^{2\pi i\f{k}{n}}(-i\ep+t-l)^{\f{1}{n}}=e^{2\pi i\f{k+1/2}{n}}(l-t+i\ep)^{\f{1}{n}},
\no
&& \bar{z}_{2k+1}\!=e^{-2\pi i\f{k}{n}}(-i\ep-t-l)^{\f{1}{n}}\!=e^{-2\pi i\f{k+1/2}{n}}(l+t+i\ep)^{\f{1}{n}} \no
&& \bar{z}_{2k+2}\!=e^{-2\pi i\f{k}{n}}(i\ep-t-l)^{\f{1}{n}}\!=e^{-2\pi i\f{k+1/2}{n}}(l+t-i\ep)^{\f{1}{n}}.\ \ \ \ \ \ \
\ea

Then we get
\ba
&& \f{\la {\mathcal O}^{\dagger}_a(w_1,\bar{w}_1){\mathcal O}_a(w_2,\bar{w}_2)\ddd {\mathcal O}_a(w_{2n},\bar{w}_{2n})\lb_{\Sigma_n}}{\left(\la {\mathcal O}_a(w_1,\bar{w}_1)^{\dagger}{\mathcal O}_a(w_2,\bar{w}_2)\lb_{\Sigma_1}\right)^n} \no
&&= {\mathcal C}_n\cdot \la {\mathcal O}^{\dagger}_a(z_1,\bar{z}_1){\mathcal O}_a(z_2,\bar{z}_2)\ddd {\mathcal O}_a(z_{2n},\bar{z}_{2n})\lb_{\Sigma_1}, \label{nrat}
\ea
where we defined
\be
{\mathcal C}_n=\left(\f{4\ep^2}{n^2(l^2-t^2)}\right)^{2n\Delta_a}\cdot \prod_{i=1}^{2n}(z_i\bar{z}_i)^{\Delta_a}.
\ee

At early time $0<t<l$, we obtain in the $\ep\to 0$ limit
\ba
&& z_{2k+1}-z_{2k+2}\simeq -\f{2i \ep}{n(l-t)}z_{2k+1}=-\f{2i \ep}{n(l-t)}z_{2k+2}, \no
&& \bar{z}_{2k+1}- \bar{z}_{2k+2}\simeq \f{2i \ep}{n(l+t)}\bar{z}_{2k+1}=\f{2i \ep}{n(l+t)}\bar{z}_{2k+2}.
\label{ean}
\ea
In this limit the $2n$ point function is factorized as follows
\ba
&& \la {\mathcal O}^{\dagger}_a(z_1,\bar{z}_1){\mathcal O}_a(z_2,\bar{z}_2)\ddd {\mathcal O}_a(z_{2n},\bar{z}_{2n})\lb_{\Sigma_1} \no
&& \simeq \prod_{k=0}^{n-1}
\la {\mathcal O}^{\dagger}_a(z_{2k+1},\bar{z}_{2k+1}){\mathcal O}_a(z_{2k+2},\bar{z}_{2k+2})\lb_{\Sigma_1} .
\label{factr}
\ea
Therefore we can confirm that the ratio (\ref{nrat}) becomes unity and this leads to
$\Delta S^{(n)}_A=0$.

\begin{figure}[httt]
  \begin{center}
   \includegraphics[width=80mm]{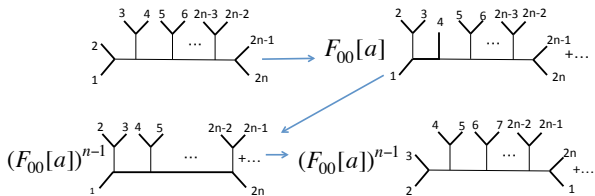}
  \end{center}
  \caption{The fusion transformations to obtain $\Delta S^{(n)}_A$.}
  \label{fig:Fmat}
\end{figure}

 On the other hand, at late time $t>l$, we find
\ba
&& z_{2k+1}-z_{2k}\simeq -\f{2i \ep}{n(l-t)}z_{2k+1}=-\f{2i \ep}{n(l-t)}z_{2k}, \no
&& \bar{z}_{2k+1}- \bar{z}_{2k+2}\simeq \f{2i \ep}{n(t+l)}\bar{z}_{2k+1}=\f{2i \ep}{n(t+l)}\bar{z}_{2k+2}.\label{lan}
\ea
In order to factorize the $2n$ point functions into $n$ two point functions, we need to
rearrange the order of the holomorphic coordinates: $[z_{1},z_2,\ddd, z_{2n}]$ in the right hand side of (\ref{factr})
as follows
\be
(z_1,z_2)(z_3,z_4)\ddd (z_{2n-1},z_{2n})\to (z_3,z_2)(z_5,z_4)\ddd (z_{1},z_{2n}).\nonumber
\ee
If we decompose this procedure into boot-strap transformations of four point functions, we can easily find that it is realized by acting $n-1$ times the fusion transformation (\ref{Ftr}) as in
Fig.\ref{fig:Fmat}. Thus we obtain
\ba
&& \la {\mathcal O}_a(z_1,\bar{z}_1){\mathcal O}_a(z_2,\bar{z}_2)\ddd {\mathcal O}_a(z_{2n},\bar{z}_{2n})\lb_{\Sigma_1}  \no
&& \simeq  (F_{00}[a])^{n-1}\cdot \left[\prod_{k=0}^{n-1}(z_{2k+1}-z_{2k})(\bar{z}_{2k+1}-\bar{z}_{2k+2})\right]^{-2\Delta_a}.
\nonumber
\ea
Finally, the ratio (\ref{nrat}) at late time is computed to be $(F_{00}[a])^{n-1}=(d_a)^{1-n}$.
In this way, we obtain the following simple formula:
\be
\Delta S^{(n)}_{A}=\log d_a. \label{finf}
\ee

Note that for any given $a$,
the quantum dimension $d_a$ is known to be the largest eigenvalue of the fusion matrix
$(N_a)^c_b(=N_{ab}^c)$. The number of primary fields contained in the operator product of $[{\mathcal O}_a]^k$ is estimated as $\sim(d_a)^k$ when $k$ is very large (see e.g. \cite{FFN}). Therefore it should be interpreted as an effective degrees of freedom included in the operator ${\mathcal O}_a$ and our results give a clear manifestation of this statement using the (Renyi) entanglement entropy.

For example, in the Ising model (i.e.(4,3) minimal model), there are three primary operators: the identity $I$, the spin $\sigma$ and the energy operator $\psi$.  Since the quantum dimension is zero for $I$ and $\psi$, $\Delta S^{(n)}_A$ is always vanishing for these. However, for the spin operator $\sigma$, we find $\Delta S^{(n)}_A=\log \s{2}$ for any $n$ as $d_{\sigma}=2$. This fact can be explicitly confirmed by using the identity \cite{Ising}:
\ba\label{Isingcorrelation}
&& \left(\la \sigma(z_1,\bar{z}_1)\sigma(z_2,\bar{z}_2)\ddd\sigma(z_{2n},\bar{z}_{2n})\lb_{\Sigma_1}\right)^2 \no
&& =\la {\mathcal O}_2(z_1,\bar{z}_1){\mathcal O}_2(z_2,\bar{z}_2)\ddd {\mathcal O}_2(z_{2n},\bar{z}_{2n})\lb_{\Sigma_1} ,
\ea
where ${\mathcal O}_2$ was defined in (\ref{opew}).

\section{Conclusions}

In conclusion, we derived the simple formula (\ref{finf}) for both Renyi ($n\geq 2$) and von-Neumann ($n=1$) entanglement entropy at late time. The essence of our proof was that the time evolution performs the fusion transformation only in left-moving sector. More generally if we consider a product of primary operators $\prod_{a}({\mathcal O}_a)^{n_a}$, we obtain $\Delta S^{(n)}_{A}=\sum_{a}n_a\log d_a$, using the sum rule in \cite{Nozaki}.
Note that the quantum dimension $d_a$ satisfies $d_ad_b=\sum_{c}N_{ab}^cd_c$.

It is interesting to note that the topological entanglement entropy \cite{wen} also has the same contribution $\log d_a$ in the presence of anyons. However, our result shows that this contribution arises in an explicit dynamical systems defined by two dimensional rational CFTs, where the consideration of time evolution played an important role. Finally it will also be an intriguing future problem to understand our results from the viewpoint of holographic entanglement entropy \cite{RT}.\\

{\bf Acknowledgements}

We thank Masahiro Nozaki for stimulating correspondences and discussions. We are grateful to Pawel Caputa and Xiao-Liang Qi for useful comments on the draft of this paper.
 SH is supported by JSPS postdoctoral fellowship for foreign researchers and by the National Natural Science Foundation of China (No.11305235). TT is supported by JSPS Grant-in-Aid for Scientific
Research (B) No.25287058 and JSPS Grant-in-Aid for Challenging
Exploratory Research No.24654057. TT is also supported by World Premier
International Research Center Initiative (WPI Initiative) from the
Japan Ministry of Education, Culture, Sports, Science and Technology
(MEXT).

\newpage
\appendix

\end{document}